\documentclass{article}
\usepackage[final]{neurips_2025} 
\usepackage[utf8]{inputenc}
\usepackage[T1]{fontenc}   
\usepackage{hyperref}     
\usepackage{url}          
\usepackage{amsfonts}    
\usepackage{nicefrac}     
\usepackage{microtype}
\usepackage{graphicx}
\usepackage{subfigure}
\usepackage{booktabs} 
\usepackage{amsmath, amsthm, amssymb}
\usepackage[capitalize,noabbrev]{cleveref}
\usepackage{tcolorbox}
\usepackage{soul}
\usepackage{xcolor}
\usepackage{bbold}
	
\usepackage{bm} 
\renewcommand{\*}[1]{\bm{#1}} 

\title{Multimodal Generative Flows for LHC Jets}

\author{%
  Darius A. Faroughy \\
  NHETC, Rutgers University \\
  \And
  Manfred Opper \\
  TU Berlin\\
\And
  C\'esar Ojeda \\
  University of Potsdam\\
 }

\begin{document}
\maketitle

\begin{abstract}
Generative modeling of high-energy collisions at the Large Hadron Collider (LHC) offers a data-driven route to simulations, anomaly detection, among other applications. A central challenge lies in the hybrid nature of \emph{particle-cloud} data: each particle carries continuous kinematic features and discrete quantum numbers such as charge and flavor. We introduce a transformer-based \emph{multimodal flow}\footnote{code repository \href{https://github.com/dfaroughy/Multimodal-flows}{\tt github.com/dfaroughy/Multimodal-flows}}  that extends flow-matching with a continuous-time Markov jump bridge to jointly model LHC jets with both modalities. Trained on CMS Open Data, our model can generate high fidelity jets with realistic kinematics, jet substructure and flavor composition.
\end{abstract}

\section{Introduction}

The Large Hadron Collider (LHC) at CERN produces billions of proton--proton collisions per second, reconstructed into final-state particles by multi-layered detectors. Among the many objects emerging from hadronic collisions, {\it jets}---collimated sprays of localized high-energy particles---play a central role in both QCD studies and searches for new physics. As {\it particle clouds}, jets have become a testbed for modern generative models. For example, they can be used to learn background distributions directly from data for resonant anomaly detection, bypassing the limitations of imperfect Monte Carlo simulations and avoiding reliance on high-level, hand-crafted observables [\cite{Buhmann:2023acn}]. These models can also act as tractable high-dimensional density estimators, offering a principled framework for evaluating how closely jet classifiers approach the theoretical optimum [\cite{geuskens2024fundamental}].

Dynamics-based frameworks such as diffusion [\cite{sohl2015deep,song2020score}] and flow-matching [\cite{albergo2022building,lipman2022flow}] have recently set new benchmarks for jet generation and now underpin state-of-the-art foundation models for particle physics [\cite{Mikuni:2024qsr}]. However, these methods operate solely on continuous spaces. This is inadequate for jets, where each particle carries both continuous kinematics and categorical attributes like charge and flavor. The use of de-quantization methods or modeling these modalities separately risks distorting physically meaningful correlations. Accurate modeling therefore requires a framework that jointly treats both continuous and discrete modalities. Multimodal flows have recently gained attention in other scientific fields [\cite{campbell2024generative}], see related work in App~\ref{app:related}.

We propose a multimodal generative framework that integrates continuous flow-matching with a continuous-time Markov jump process for the discrete dynamics. This yields a unified probabilistic path over hybrid spaces capable of generating both kinematics and quantum numbers for jet constituents. Trained on CMS Open Data, our model accurately reproduces the kinematics, substructure and flavor content of real world jets. We show that our model, when equipped with a multimodal particle transformer architecture, with mode-specific and fused encoders can produce state of the art results on the \textsc{AspenOpenJets} dataset introduced by \cite{Amram:2024fjg}.

\section{CMS Open Data}\label{sec:aoj_dataset}  

\begin{figure*}[t]
    \centering   
   \includegraphics[width=1\linewidth]{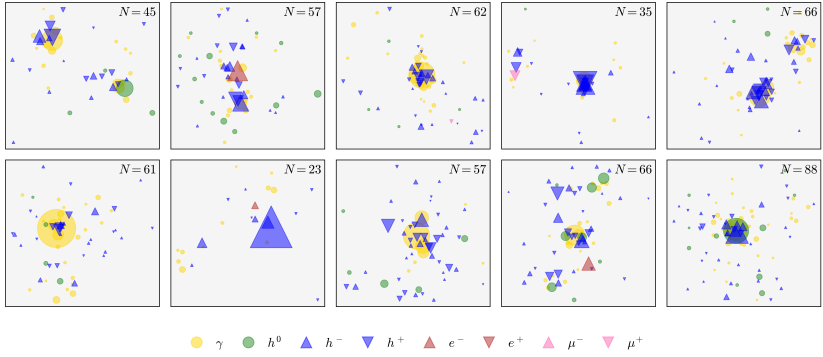}
    \caption{Visualizations of 10 CMS jets in the $(\eta,\phi)$-plane taken from the {\sc AspenOpenJets} dataset. Each particle cloud is centered around the jet axis. The size of each marker is proportional to the particle's transverse momentum ($p_T$), while the shape and color encode electric charge and flavor, respectively. The upper right corner of each panel indicates the total number of constituents in the respective jet.}
    \label{fig:particle_clouds}
\end{figure*}

In this work we are interested in training generative models on real LHC jets. We use the recently released \textsc{AspenOpenJets} (AOJ) dataset derived from 13~TeV proton-proton collisions recorded by CMS in 2016. Each jet is represented as a particle-cloud $\mathbf{z}=\{z^d\}_{d=1}^{D}$ with up to $D=150$ constituents, where each particle is described by (continuous) kinematic features in hadronic coordinates and a (categorical) {\it flavor} token: $z^d\equiv(x^d,\,k^d)\in \mathbb{R}^3\otimes\mathcal{F}$ with $x^d\in (p_T,\, \Delta\eta,\, \Delta\phi)$ and $k^d\in \mathcal{F}=\{\gamma,\, h^0,\, h^-,\,h^+,\, e^-,\, e^+,\, \mu^-,\,\mu^+ \}$. Here $p_T=\sqrt{p_x^2+p_y^2}$ is the transverse momentum, and $\Delta\eta$, $\Delta\phi$ are coordinates relative to the jet axis. The possible flavors are: photons ($\gamma$), electrically charged hadrons ($h^-\,,\ h^+$), electrically neutral hadrons ($h^0$), electrons ($e^-$), positrons ($e^+$), muons $(\mu^-)$ and anti-muons ($\mu^+$), corresponding to particle species and charges reconstructed by the CMS tracking system and calorimeters. Representative particle clouds from the AOJ dataset are shown in Fig.~\ref{fig:particle_clouds}. The flavor composition is strongly imbalanced: photons and charged hadrons make up $\sim$90\% of all constituents (in almost equal proportions), neutral hadrons $\sim$10\%, while leptons occur only at the per-mille level.

\section{Multimodal flows for particle clouds}\label{sec:multimodal}

We now describe our extension of flow-matching over the hybrid space $\mathbb{R}^3\otimes\mathcal{F}$. We provide extensive supplementary material to this section in App.~\ref{sec:generative_modeling}. We denote by $\*z_t=(\*x_t,\*k_t)$, the time-dependent path of the jet that transforms an arbitrary source data $\*z_0=(\*x_0,\*k_0)\sim\mu$ at $t=0$ into the target data $\*z_1=(\*x_1, \*k_1)\sim\nu$ at $t=1$ and $P_t(\*z)$ the corresponding probability path satisfying

\begin{align}\label{eq:multimodal_prob_flow}
  \partial_t {P}_t(\*z_t) = - \nabla_{\*x} \cdot\left[{\*u}_t(\*z_t)\, {P}_t(\*z_t)\right]\ +\ \sum_{\*j \neq \*k_t} \left[ {\*W}_t(\*z_t,\*j) \, {P}_t(\*j) \ -\  {\*W}_t(\*j,\*z_t) \, {P}_t(\*z_t)\right] \,,
\end{align}

and subject this the boundary conditions $P_0 = \mu(\mathbf{x}_0, \mathbf{k}_0)$ and $P_1 = \nu(\mathbf{x}_1, \mathbf{k}_1)$. The first term is the familiar continuity equation for continuous densities where $\*u_t$ is the velocity field acting on the particle kinematics $\*x\in\mathbb{R}^3$, while the remaining terms describe a continuous time Markov jump bridge generating discrete transitions between flavor tokens $\*k\in\mathcal{F}$ with a jump rate matrix $\*W_t$. Following the conditional flow-matching framework, we introduce conditional quantities and write the path as a marginalization over the boundary data:
\begin{align}\label{eq:decompose_hybrid}
P_t(\mathbf{z}_t) =  
\sum_{\mathbf{k}_0, \mathbf{k}_1}\!\! \int \!\!{\mathrm{d}}\mathbf{x}_0 {\mathrm{d}}\mathbf{x}_1  \,\mu(\*x_0,\*k_0)\,\nu(\*x_1,\*k_1)\,
\prod_{d=1}^D & p_t(x_t^d | x^d_0, x^d_1) \,
 q_t(k_t^d | k^d_0, k^d_1).
\end{align}
Here, to ensure tractability we impose a complete factorization over particles and both modalities, where the continuous probability density $p_t(x^d | x^d_0, x^d_1)$ and discrete probability mass function $q_t(k^d | k^d_0, k^d_1)$, satisfy separate conditional dynamics. Application of this marginalization trick yields a velocity field ${\*u}_t$ and a jump rate ${\*W}_t$ expressed in terms of the expectations of the conditional velocities and rates with respect to the {\it posterior probability} distributions:
\begin{align}\label{eq:marginal_vec}
    \*u_t(\*z_t) & = \mathbb{E}_{\pi_t(\*z_0,\*z_1|\*z_t)} \, \*u_t(\*x_t|\*x_0,\*x_1)\\ 
    \*W_t(\*j, \*z_t) &= \mathbb{E}_{\pi_t(\*z_0,\*z_1|\*z_t)} \, \*W_t(\*k_t,\*j|\*k_0,\*k_1)\,,\label{eq:marginal_rate}
\end{align}

where the posterior follows 

\begin{equation}
\pi_t(\*z_0, \*z_1|\*z) = {p}_t(\*x | \*x_0, \*x_1) {q}_t(\*k | \*k_0, \*k_1) \mu( \*z_0) \nu(\*z_1)\, /\, P_t(\*z)
\end{equation}

from Bayes theorem. Although each component evolves independently under the conditional process, non-trivial correlations between particles and their respective modalities emerge for the generative process \eqref{eq:decompose_hybrid}.

\begin{figure}[t!]
    \centering
    \includegraphics[width=1\linewidth]{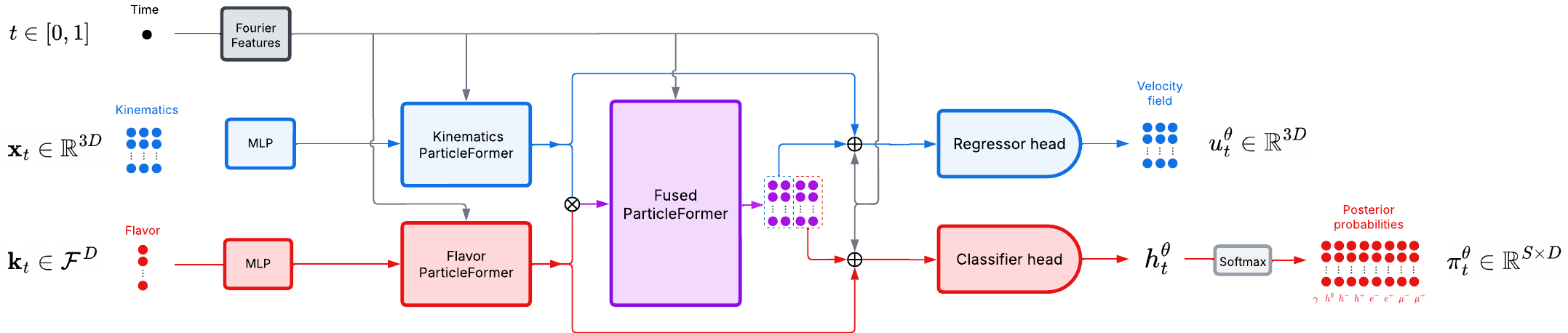}
    \caption{Multi-modal particle transformer architecture. Additional details are provided App.~\ref{app:architecture}.}\label{fig:architecture}
\end{figure}

\paragraph{Conditional dynamics} The next step is to specify the conditional dynamics over $\mathbb{R}^3\otimes \mathcal{F}$. For the continuous modality, the standard choice is to take a {\it uniform flow} that transports source data into target data through straight paths with constant velocities [\cite{liu2022flowstraightfastlearning}]. For the discrete modality we propose a multi-state generalization of the {\it random telegraph process} [\cite{gardiner2010stochastic}], a continuous-time Markov process originally used to model noise in binary communication channels (bit-flips). As shown in App.~\ref{sec:jumps}, the resulting velocity field and jump rate matrix are given by:

\begin{align}
   & u^d_t(x_t^d|x^d_0,x^d_1) = x^d_1-x^d_0\\ 
   & W^d_t(k^d_t=i,j^d=j|k^d_1=k) = 1 + \frac{ S\,\omega_{t}}{1 - \omega_{t}}\, \delta_{ik} + \omega_{t} \,\delta_{jk} .\label{eq:sol_MJB}
\end{align}

where $S$ is the token vocabulary size, $\omega_{t}\equiv \exp(-S\beta(1-t))$ and $\beta$ is a stochasticity hyperparameter that controls the frequency of jumps per unit time. 

\paragraph{Multimodal objective} Substituting these solutions into Eqs.~\eqref{eq:marginal_vec} 
and~\eqref{eq:marginal_rate}, the expectation of the velocity field cannot be expressed in closed form and must be approximated from data with the conditional flow-matching objective by regressing the conditional vector field with a parametric function $u_t^\theta(\*z_t)$. By contrast, the expectation over the discrete jump rates is fully tractable and yields the explicit expression
\begin{align}
W_t^d(k^d=i,\,j^d=j) = 1 + \frac{S\,\omega_{t}}{1-\omega_{t}}\,\pi_t(k_1^d=i\,|\,j^d=j) + \omega_{t}\,\pi_t(k_1^d=j\,|\,j^d=j),
\label{eq:target_rate}
\end{align}
where $\pi_t$ denotes the posterior distribution for particle $d$ over the final state token $i$ conditioned on the intermediate state $j$ at time $t$. Therefore, if the posterior $\pi_t$ can be approximated with a parametric estimate $\pi_t^\theta$ from the data, one can directly compute the rates via \eqref{eq:target_rate}. In this case the posterior learning problem is equivalent to a multi--class classification task. We thus introduce a time-dependent classifier $h^\theta_t$ such that the vector of posterior probabilities is given by the \textit{softmax} function $\pi^\theta_t = {\tt softmax}(h_t^\theta)$ and train the classifier function by minimizing the cross-entropy loss. 

We train using the flow-matching mean-square error (MSE) loss for the particle kinematics and the cross-entropy (CE) loss for the particle flavor tokens. Following the approach of \cite{kendall2018multi}, these two objectives are combined into a single weighted loss:

\begin{equation}\label{eq:multimodal_loss}
    \mathcal{L}_{\rm MMF} \ =\ \mathbb{E}_{t,\,(\*z_0,\*z_1),\,\*z_t} \left[\frac{|| u^\theta_t(\*z_t)- u_t(\*x_t|\*x_0,\*x_1)||^2}{2(\sigma^1_t)^2}\, -\ \frac{\log h_t^\theta(\*z_t, \*k_1)}{2(\sigma^1_t)^2} \ +\ \log\!\left(\sigma^1_t\,\sigma_t^2\right) \right].
\end{equation}  

In the expectation, time is drawn uniformly $t\sim\mathcal{U}[0,1]$ over the unit interval, pairs of source and target points drawn from the coupling $(\*z_0,\*z_1)\sim\mu\otimes\nu$ and $\*z_t\sim P_t(\cdot|\*z_0,\*z_1)$. In contrast to \cite{kendall2018multi}, where the uncertainty weights $\sigma_i$ are fixed trainable scalars, we promote them to time-dependent functions $\sigma^i_t$. In practice, we parametrize the weights via $\sigma_t^i=\exp(-w_t^i)$, where $w_t$ is the output of an {\it uncertainty network} that we discard during inference. This allows the relative weighting between modalities to adapt dynamically along the generative path, enabling the model to prioritize different objectives at different stages of the evolution. Besides improving the convergence of the training, this formulation eliminates the need for manually tuning loss weights.

\paragraph{Multimodal ParticleFormer}\label{sec:architecture} To learn the continuous and discrete modalities together, we approximate the conditional velocity field and the posterior classifier function with a single permutation equivariant neural network $u^\theta_t\otimes h^\theta_t$. The overall architecture, depicted in Fig.~\ref{fig:architecture}, has three components: 1) two mode-specific encoders, 2) a {\it fused} encoder, and 3) two task-specific heads. All three encoders consist of non-causal particle transformers [\cite{Qu:2022mxj}] with stacked multi-head self-attention blocks. The {\it regressor head} predicts the continuous-valued vector field $u^\theta_t$ for the MSE loss, while the {\it classifier head} outputs the logits $h^\theta_t$ for the CE loss. Fore more details on the architecture see App.~\ref{app:architecture}.

\paragraph{Sampling} Once the model is trained, new samples can be generated from a source input $\*z_0$ by simulating the associated dynamics to the probability flow equation \eqref{eq:multimodal_prob_flow}. This entails numerically solving the joint dynamics for both the continuous and discrete components. For the continuous dynamics we solve the ordinary differential equation $\dot{\*x}_t=\*u_t^\theta(\*x_t)$ with Euler's method using discrete time-steps of size $\Delta t$. The discrete dynamics of the flavor tokens follows the multivariate telegraph process governed by the rate $\mathbf{W}^\theta_t$ derived from the trained posterior $\pi_t^\theta={\tt softmax}(h^\theta_t/T)$. Here we have introduced a temperature scaling hyperparameter $T$ [\cite{guo2017calibration}] that helps improving sampling quality. To efficiently simulate this stochastic process, we employ \textit{$\tau$-leaping} [\cite{gillespie2001approximate, NEURIPS2022_b5b52876}], a well-established approximation method widely used in chemical reaction kinetics. More details on the sampling algorithm can be found in App.~\ref{app:sampling}. 

\section{Experiments}
\label{sec:results}

\begin{figure}[t]
    \centering
    \includegraphics[width=1.0\linewidth]{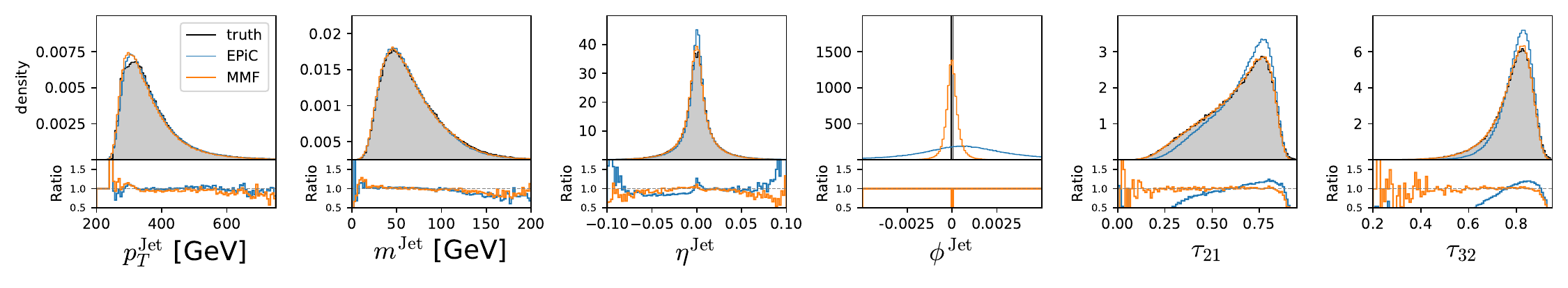}\\
    \includegraphics[width=1.0\linewidth]{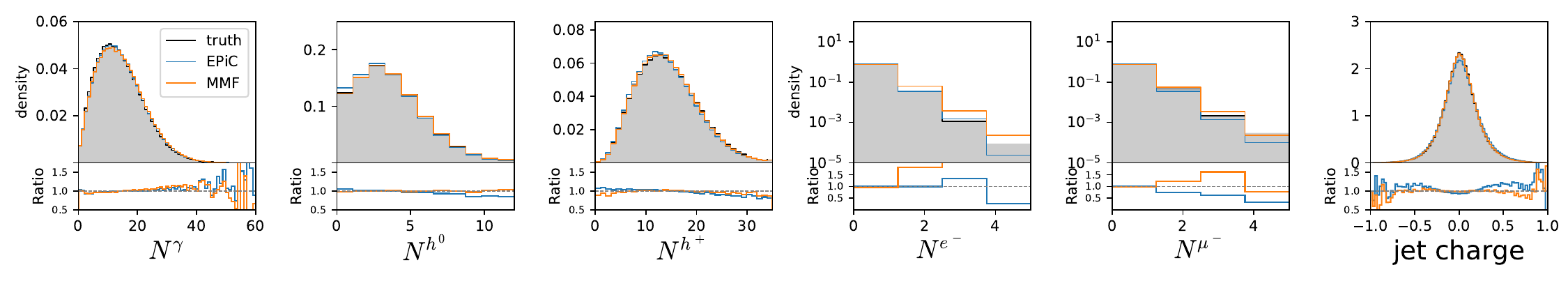}
    \caption{Performance comparison between generated samples from our particle transformer MMF model (orange) and the EPiC-FM baseline (blue) for various high-level jet observables. The corresponding Wasserstein distance between the generated and test distributions are shown in Table \ref{tab:results}.} \label{fig:res}
\end{figure}

\paragraph{Aspen Open Jets} We demonstrate that our transformer-based multimodal flow (MMF) is capable of generating the particle kinematics and flavor composition of real-world jets from CMS open data. Since our work consists of a proof-of-concept, we do not attempt a full optimization of the architectures and only train our models on a subset of the {\tt AspenOpenJets} dataset. Our training dataset consists of $1.25$ million AOJ jets, with $1$ million jets for training and $250$K jets for validation. For the source data $\*z_0=\{z_0^d\}_{d=1}^{D}$, we draw point-cloud noise from the distribution $z^d_0=(x^d_0,\,k^d_0) \sim  \nu\equiv\mathcal{N}(0, \mathbf{1})\,\otimes\, {\mathcal U}(p)$ where $\mathcal{N}$ denotes the standard Gaussian over $\mathbb{R}^3$ and $\mathcal{U}$ represents the uniform categorical distribution over the token space $\mathcal{F}$ with probability parameter $p=1/S$. We compare our proposed model to EPiC-FM, a permutation equivariant flow-matching model with deep-sets architecture designed for particle cloud data that achieves state-of-the-art results on simulated jets [\cite{Buhmann:2023pmh,Buhmann:2023zgc,Birk:2023efj}]. Our training setup is described in App.~\ref{app:experiments}. We generate samples with 270K jets using a temperature scaling of $T=0.85$ and solve the dynamics with $\Delta t=0.001$ time-step size. To quantify the performance of the continuous component of our generator we compute the jet-level kinematics $(p_T, m, \eta, \phi)$ and the $N$-subjettiness ratios $\tau_{21}$, $\tau_{32}$ [\cite{Thaler:2010tr}] as probes for jet substructure. For the discrete component we examine the particle flavor multiplicities $N^{k}$, $k\in{\mathcal{F}}$. Finally, to test the model capacity in reproducing cross-modal correlations we compute the {\it jet charge} 
\begin{equation}
\mathcal{Q}= \frac{1}{p_T^{\rm jet}}\sum_{i\in\rm jet} Q_i\,{{p_T}_i},
\end{equation}
a non-trivial aggregate of the particles electric charge $Q_i$ and transverse momentum ${p_T}_i$. These distributions, and the corresponding Wasserstein-1 distances to the AOJ truth, are provided in Fig.~\ref{fig:res} and Table~\ref{tab:results}, respectively. 

\begin{table}[t]
  \caption{The Wasserstein distances $W_1^{\mathcal{O}}$ computed between the generated data and the test data for each jet observables $\mathcal{O}$. Lower values are better. }
  \label{sample-table}
  \resizebox{\columnwidth}{!}{%
  \begin{tabular}{lccccccc}
    \toprule
    Model & $W_1^{p_T}$ & $W_1^{m}$ & $W_1^{\eta}$ & $W_1^{\phi}$ & $W_1^{\tau_{21}}$ & $W_1^{\tau{32}}$ & $W_1^{\mathcal{Q}}$\\
    \midrule
    EPiC-FM    & {$\bf 0.92$} & $1.63$ & $1.2\times 10^{-3}$ & $2.8\times 10^{-3}$ & $3.1\times 10^{-2}$  & $1.8\times 10^{-2}$ & $9.5\times 10^{-3}$  \\
    MMF (ours) & $4.64$ & ${\bf1.26}$ & ${\bf 6.3\times 10^{-4}}$ & ${\bf 2.3\times 10^{-4}}$ & ${\bf 2.3\times 10^{-3}}$ & ${\bf 2.8\times 10^{-3}}$ & ${\bf 1.4\times 10^{-3}}$     \\
    \bottomrule
    \\
  \end{tabular}
  }
    \resizebox{\columnwidth}{!}{%
    \begin{tabular}{lcccccccc}
    \toprule
    Model & $W_1^{N^{\gamma}}$ & $W_1^{N^{h^0}}$ & $W_1^{N^{h^-}}$ & $W_1^{N^{h^+}}$ & $W_1^{N^{e^-}}$ & $W_1^{N^{e^+}}$ & $W_1^{N^{\mu^-}}$ & $W_1^{N^{\mu^+}}$ \\
    \midrule
    EPiC-FM    & $\bf 0.23$ & $0.10$& $0.28$ & $0.23$& $\bf5.6\times 10^{-4}$ &  $\bf6.8\times 10^{-4}$ & $\bf2.6\times 10^{-3}$ & $\bf3.2\times 10^{-3}$ \\
    MMF (ours) & $0.34$ & $\bf 0.01$ & $\bf0.09$ & $\bf0.10$ & $5.7\times 10^{-2}$ & $5.6\times 10^{-2}$ & $4.3\times 10^{-2}$ & $4.3\times 10^{-2}$ \\
    \bottomrule
  \end{tabular}\label{tab:results}
  }
\end{table}

\paragraph{Discussion}  
The results show that our MMF model outperforms the EPiC-FM baseline in several respects: (i) jets are more accurately centered in the $\eta$--$\phi$ plane, (ii) jet substructure observables are reproduced with higher fidelity, and (iii) jet charge distributions agree more closely with data, indicating superior modeling of cross-modal correlations. We attribute these improvements to the particle transformer architecture, which better captures inter-particle correlations than the Deep Sets--based EPiC encoder\footnote{Interestingly, both methods have some trouble capturing the irregular shape of the $p_T$ peak, with the MMF model performing slightly worst than EPiC-FM.}. On the other hand, EPiC-FM provides good modeling of all flavor multiplicities, successfully reproducing dominant ($\gamma,\,h^\pm$), subdominant ($h^0$), and even rare ($e^\pm,\mu^\pm$) classes. MMF outperforms EPiC-FM for the dominant and subdominant flavors but slightly underperforms for the rare leptons, whose per-mille frequency in the training data remains a limiting factor. We hypothesize that residual stochastic noise from the $\tau$-leaping sampler affects these minority classes near the time endpoint $t\approx1$. If true, this limitation could be mitigated by applying a post-sampling calibration to the generated jets. Finally, we note that the choice of temperature scaling is critical: while $T=0.85$ yields unbiased multiplicities, when scanning over $T$ during our experiments, we found that departures from this value systematically distort the neutral-hadron distribution $N^{h^0}$. This highlights the importance of this hyperparameter in our setup.

\begin{ack}
Darius Faroughy was supported by DOE grant DOE-SC0010008. C\'esar Ojeda was supported by the Deutsche Forschungsgemeinschaft (DFG)– Project-ID 318763901– SFB1294. This research used resources of the National Energy Research Scientific Computing Center, a DOE Office of Science User Facility supported by the Office of Science of the U.S. Department of Energy under Contract No. DE-AC02-05CH11231 using NERSC award HEP-ERCAP0027491
\end{ack}

\bibliography{neurips_2025}
\bibliographystyle{neurips_2025}
\newpage
\appendix
\section*{Supplementary Material}
\section{Related Work}
\label{app:related}
Generative models that independently handle discrete and continuous variables have been explored since the inception of research in both denoising and diffusion frameworks, with various methodologies attempting to relate these approaches. \cite{winkler2024bridging} show that a limit for Ehrenfest processes defined on discrete spaces converges to an Ornstein-Uhlenbeck dynamic. Further extending the theory, \cite{ren2024discrete} establish a stochastic integral formulation of discrete diffusion models, generalizing the Poisson random measure through a L\'evy-type integral. The prevalence of multimodal flows in the natural sciences has spurred additional multimodal solutions akin to our work [\cite{lin2025tfg}]. Purely discrete flows were developed by \cite{gat2024discrete}, while in protein design, \cite{campbell2024generative} introduced a fully multimodal methodology. Their approach differs from ours in selecting conditional paths in an ad hoc manner, directly interpolating within probability space. Conversely, we introduce a more general strategy allowing diverse probabilistic paths by leveraging Markov bridges, as presented by \cite{chopin2023computational}. Finally, a general extension of flow matching theory to a broader family of stochastic process generators was proposed by \cite{holderrieth2024generator,lipman2024flow}. Our method can be viewed as a special instance of this broader framework, where the probability path is derived via a Markov bridge applied to a specific type of reference process.

\section{Generative modeling with bridge processes}
\label{sec:generative_modeling}
In dynamic generative modeling, one has access to samples from a target distribution $\nu$, and the goal is to generate novel unseen samples from that same distribution. To generate new samples, we start with a tractable source distribution $\mu$ (e.g., a Gaussian) and transform its samples through a deterministic or stochastic generative process $\{\*z_t\}_{t\in[0,1]}$, such that by the final time $t=1$, the distribution  of the transformed variables $\*z_1$ match the desired distribution $\nu$. Any such transformation has an associated probability path $P_t(\*z)$, and the objective is to construct the transformation such that the probability path ensures $P_0 = \mu$ and $P_1 = \nu$. In the following, we will refer to the probability path that complies to this boundary conditions as the \textit{target probability path} \footnote{Contrary to the diffusion literature, where the generative process is the backward process running in reverse-time, we follow throughout this paper the flow-based convention where the generative process is forward-time.}.

Stochastic transformations can be specified as stochastic processes through an infinitesimal generator $\mathcal{L}_t$ or its adjoint operator and the corresponding Fokker--Planck or Master equations, whereas deterministic transformations can be achieved with a \textit{flow} $\psi_t$ specified in turn with a velocity field $u_t$. The family of dynamic generative models, constitute a group of methodologies that are able to construct these transformations via neural networks approximations to the desired target generators $\mathcal{L}^\theta_t$ or vector fields $u^\theta_t$. In particular, the flow matching family achieves this through a conditional strategy, whereby one designs a conditional probability path that, after marginalization, recovers the target path. Crucially, the conditional path is defined to ensure access to tractable conditional generators or velocities. The key insight of this methodology is that the target generator or velocity field can be constructed or learned by properly averaging the conditional generators. The average however, is performed over a posterior distribution from the conditional paths.

In our formulation, we closely follow the flow matching methodology of \cite{albergo2022building,lipman2022flow} and start by constructing probability paths that follow a prescribed continuity equation.  We then construct conditional probability paths with the help of \textit{reference process} with known generators that enables the construction of \textit{bridges} between point samples, ensuring transformations from $\*z_0$ to $\*z_1$.  In  the context of particle clouds for jets, we will assume that  $\*z = ( \*x,  \*k)$ is a collection of continuous ($\*x$) and discrete ($\*k$) vectorial random variables.  And in the following we will show the construction for $\*x$ and $\*k$ separately.

\subsection{Continuous random variables}\label{sec:FM}

First, we consider the continuous case with $\*z \equiv \*x$ and $P_t \equiv p_t$. Our goal is to construct probability paths that satisfy the continuity equation:  
\begin{equation}
  \partial_t p_t(\*x) = -\nabla \cdot \big[ \*u_t(\*x) \, p_t(\*x) \big] \,
  \label{eq:continuity_equation}
\end{equation}
subject to the boundary conditions $p_1 = \nu$ and $p_0 = \mu$. It is known that such continuity equations specify the transformation of the random variables $X_0 \sim \mu$ with a flow $\*x_t = \psi_t(\*x_0)$ whose dynamics are given by: 
\begin{equation}\label{eq:ode}
   \frac{\mathrm{d} \*\psi_t(\*x)}{\mathrm{d}t} =  u_t(\psi_t(\*x))  \ ,\;\mbox{with}\;  \psi_{t=0}(\*x) =  \*x .
\end{equation}
Where $\*u_t$ is the target velocity field. Using the flow $\psi_t$ the marginal PDFs can be obtained via the push back formula $p_t(\*x) = [\psi_{t\#}p](\*x)$.

\paragraph{Conditional Flow-Matching}
Since one does not have access to close forms of $\*u_t$ that solve equation \eqref{eq:continuity_equation} given the boundary conditions, we will construct it by first introducing conditional probability paths $p_t(\*x | \*x_0, \*x_1)$ that are able to obtain the target probability path $p_t$ by marginalizing  over the target $\nu$ and data distribution $\mu$:
\begin{equation}\label{eq:decompose}
p_t(\*x) = \int  p_t(\*x | \*x_0, \*x_1) \mu(\*x_0) \nu(\*x_1) d\*x_0  d\*x_1\,
\end{equation}
the desired boundary conditions of the target path are obtained if the conditional fulfill:
\begin{eqnarray}\label{eq:bridge_limits}
\lim_{t\to 0} p_t(\*x | \*x_0, \*x_1) = \delta(\*x - \*x_0) \\
\lim_{t\to 1} p_t(\*x | \*x_0, \*x_1) = \delta(\*x - \*x_1)\,
\end{eqnarray}
here $\delta(\cdot)$ denotes the Dirac function. Many such conditional paths can be constructed with these boundary conditions, in the standard flow matching methodology, this is achieved by linearly interpolating between the end points: 
\begin{equation}\label{eq:interpolation}
\*x_t = t\*x_1 + (1-t)\*x_0  
\end{equation}
Hence, for this construction, we obtain {\it Dirac} probability paths ${p}_t(\*x | \*x_0, \*x_1) = \delta(\*x - \*x_t)$. 
One can show that such probability paths are generated by the following conditional vector field:
\begin{equation}
 \*u_t(\*x|\*x_0,\*x_1) = \*x_1 - \*x_0
\end{equation}
Now one can prove [\cite{albergo2022building,lipman2022flow}] that the desired target vector field, can be obtained from the conditional by:
\begin{equation}
\*u_t(\*x) =\mathbb{E}_{\rho_t(\*x_0,\*x_1|\*x)}\,\left[\*u_t(\*x|\*x_0,\*x_1)\right]\,\label{eq:marginalization_trick}
\end{equation}
where the expectation is performed with respect to the \textit{posterior probability} density of end point pairs $(\*x_0,\*x_1)$ conditioned on the intermediate point $\*x_t = \*x$  at time $t$. This posterior is obtained  by applying Bayes formula:
\begin{eqnarray}
\rho_t(\*x_0,\*x_1|\*x) =  \frac{{p}_t(\*x | \*x_0, \*x_1) \mu( \*x_0) \nu(\*x_1) }{p_t(\*x)}\,.
\end{eqnarray}
In order to compute \eqref{eq:marginalization_trick} we can use the widely known fact that the conditional expectation can be expressed as the minimizer of an appropriate least-square error problem equivalent to
a nonlinear regression problem. We use a neural network approximation $\*u_t^\theta(\*x_t)$ to the minimizer, where $\theta$ corresponds to the parameters of the network. To be precise, the conditional expectation is
obtained as the minimum of the mean square error (MSE) loss: 
\begin{equation}\label{eq:CFM}
    \mathcal{L}_{\rm CFM}= \mathbb{E}_{t,(\*x_0,\*x_1),\*x_t}\, || \*u_t^\theta(\*x_t) - \*u_t(\*x_t|\*x_0,\*x_1)||^2
\end{equation}
The expectation runs over $t\sim \mathcal{U}(0,1)$, $(\*x_0,\*x_1)\sim\mu\otimes \nu$ and $\*x_t\sim p_t(\cdot|\*x_0,\*x_1)$.

\subsection{Discrete random variables}\label{sec:jumps}

We now consider the case where the data $\*z \equiv \*k$, $P_t\equiv q_t$, takes values in a discrete space $\*k \in \{0, 1, \ldots, S\}^D$, and where the source is obtained from a simple probability mass function $\*k_0\!\sim\!\mu$ and one has, as before, access to the target distribution via data samples from $\nu$. Our goal is to  construct a target probability path $q_t(\*k)$ between $\mu(\*k)$ and $\nu(\*k)$. Similar to (\ref{eq:continuity_equation}) we start by imposing a continuity equation to the target probability path, which for the discrete variable case corresponds to the \textit{Master Equation}:
\begin{equation}\label{eq:master_equation}
\partial_t {q}_t(\*k) = \sum_{\*j \neq \*k} \big[{\*W}_t(\*k,\*j) \, {q}_t(\*j) -
 {\*W}_t(\*j, \*k) \, {q}_t(\*k)\big]\, 
\end{equation}
and subject this equation to the boundary conditions $q_0(\*k) = \mu$ and $q_1(\*k) = \nu$. This is a type of stochastic process in which transitions, or {\it jumps}, between discrete states occur at continuous random times. The process
is fully defined by  rate matrices ${\*W}_t \in \mathbb{R}^{(S+1)^D \times (S+1)^D}$, where for $\*k \neq \*j$, 
$\*W_t(\*k,\*j) \,{\rm d}t$ equals the average number of jumps  from state $\*j$ to state $\*k$ at time $t$ occurring during an infinitesimal time window ${\rm d}t$. Formally:
\begin{equation}
\*W_t(\*k,\*j) \doteq \lim_{\Delta t \to 0} \frac{q_{t + \Delta t | t}(\*k| \*j) - \delta_{\*k, \*j}}{\Delta t}\label{eq:def_rate}
\end{equation}
where we have defined the transition probability by $q_{t|s}(\cdot|\cdot) $
and the symbol $q$ denotes the computation of probabilities with respect to $q$. Since transition probabilities are normalized, one has
$\*W_t(\*k,\*k) = -\sum_{\*k'\neq \*k} \*W_t(\*k',\*k)$. 

Now, since the desired target rate ${\*W}_t$ that fulfills \eqref{eq:master_equation} subject to the boundary conditions is unknown, we proceed  as before by introducing a conditional probability path such that:
\begin{align}\label{eq:combined}
    &q_t(\*k) = \sum_{\*k_0, \*k_1} q_t(\*k | \*k_0, \*k_1) \mu(\*k_0) \nu(\*k_1), \\
    &\lim\limits_{t\to 0} q_t(\*k | \*k_0, \*k_1) = \delta_{\*k, \*k_0}, \\
    &\lim\limits_{t\to 1} q_t(\*k | \*k_0, \*k_1) = \delta_{\*k, \*k_1}.
\end{align}
where $\delta$ now correspond to the Kronecker delta function. 

\paragraph{Construction of bridge processes}
We are now poised with the task of constructing a conditional probability path with tractable rates, that achieve conditions \eqref{eq:combined}. Since we are dealing with discrete variables, direct interpolation in data space as achieved by equation \eqref{eq:interpolation} is impossible. We will follow instead  \cite{fitzsimmons1992markovian} and construct the conditional probability path as a Markov Bridge. 

We assume access to a \textit{reference process} characterized by the rate $ \*R_t(\*k, \*j) $. For this process, we also assume we have closed-form solutions to its corresponding master equation Eq.~\eqref{eq:master_equation}, expressed in terms of the conditional probability $ \tilde{q}_{s|t}(\*k_1 | \*k) $
of being in state $\*k_1$ at time $s$ when the state was $\*k$ at time $t$. Following \cite{fitzsimmons1992markovian}, we can now construct a Markov bridge that satisfies \eqref{eq:combined} with a tractable conditional rate given by:
\begin{equation}\label{eq:bridge_rate}
\*W_t(\*k,\*j|\*k_0,\*k_1) = \*R_t(\*k,\*j) \frac{\tilde{q}_{1|t}(\*k_1 | \*k)}{\tilde{q}_{1|t}(\*k_1 | \*j)}\,.
\end{equation}
Note the slight abuse of notation: the conditional rate is {\it not} a conditional probability. 
We also have a corresponding close form for the conditional distribution:
\begin{equation}\label{eq:discrete_bridge}
q_t(\*k |\*k_0,\*k_1) = \frac{\tilde{q}_{1|t}(\*k_1 | \*k) \tilde{q}_{t|0}(\*k | \*k_0)}{\tilde{q}_{1|0}(\*k_1 | \*k_0)}.
\end{equation}
Note, that the rate only depends on the final time condition $\*k_1$, but not on the initial state $\*k_0$. 
This fact is a result of the Markov nature of  the reference process. Now since ${\*W}_t(\cdot,\cdot|\*k_0,\*k_1)$ is known, we can proceed similarly to the flow matching methodology and find the desired target rate trough the marginalization trick as wshown in the next paragraph: 
\begin{equation}\label{eq:marginalization_trick_rate}
   \* W_t(\*k,\*j) = \mathbb{E}_{\pi_t(\*k_0, \*k_1|\*k)}\,\left[ {\*W}_t(\*k,\,\*j|\*k_0, \*k_1)\right]   
\end{equation}
this is a similar result to equation \eqref{eq:marginalization_trick}, where  now the rate $\*W_t(\*k,\*j)$ fulfills the Master equation equation \eqref{eq:master_equation} while attaining the target probability flow $q_t$. 
The expectation is again over the posterior probability of the end points,  now conditioned on the state $\*k$ at time $t$, obtained with Bayes theorem as:
\begin{eqnarray}\label{eq:poster_disc_discrete}
\pi_t(\*k_0, \*k_1|\*k) =  \frac{{q}_t(\*k | \*k_0, \*k_1) \mu( \*k_0) \nu(\*k_1) }{q_t(\*k)}.
\end{eqnarray}
If one is then able to obtain parametric estimate $\pi_t^\theta$ of the posterior 
 $\pi_t$ (\ref{eq:poster_disc_discrete}) from the data, one should be capable of obtaining the target rates by performing the average \eqref{eq:marginalization_trick_rate}. 
 We can view the posterior learning problem as a probabilistic multi--class classification task. We thus introduce a time-dependent neural network classifier $h^\theta_t$ such that the vector of posterior probabilities is given by the \textit{softmax} function ${\tt softmax}(h_t^\theta)$. We train the classifier function with the cross-entropy loss, which in this context will be referred to as the Markov jump bridge loss:
\begin{equation}\label{eq:cross_entropy_loss}
    \mathcal{L}_{\rm MJB} = \mathbb{E}_{t,(\*k_0,\*k_1),\*k_t}\,\log h_t^\theta(\*k_t, \*k_1)
\end{equation}
where the expectation runs over $t\sim \mathcal{U}(0,1)$, $(\*k_0,\*k_1)\sim\mu\times \nu$ and $\*k_t\sim q_t(\cdot|\*k_0,\*k_1)$.
Similar to the square loss problem (\ref{eq:CFM}), one can show that  that the minimizer of the cross entropy loss is given by the proper conditional (posterior) probability as one can relate the cross entropy loss to the regression problem [\cite{bishop2006pattern}] (it suffices to see the regression problem over the one hot encoding expression of the likelihood).  In the following we impose further conditions on our reference process such as to make our computations more tractable.

\paragraph{Factorizing over Dimensions}
In our application, we have to model $D$ dimensional data of the form $\*k \in \{0, 1, \ldots, S\}^D$.  Here the state space size grows exponentially with $D$, which  leads to computationally intractable target processes transitions.  To avoid this problem we will 
factorize first the reference process such that all coordinates $k_t^d$ of $\*k_t$ are independent processes defined by individual transition rates $R_t^d$.
\begin{eqnarray}
R_t(\*k,\*j) = \sum^D_{d=1} R_{t}^{d} (k^d , j^d)\prod_{l\neq d} \delta_{j^l,k^l}
\label{eq:trans_rate_indep}
\end{eqnarray}
where for each dimension the marginal probability will follow its own master equation:
\begin{equation}
\frac{\partial \tilde{q}^d_t(k^d)}{\partial t} = \sum_{j^d \neq k^d} R^d_{t}(k^d, j^d) \tilde{q}^d_t(j) - \sum_{j^d \neq k^d} R^d_{t}(j^d ,  k^d) \tilde{q}^d_t(k^d)
\label{eq:Chap_Kol_oned}
\end{equation}
And the full marginal follows $\tilde{q}_t(\mathbf{k}) = \prod_d \tilde{q}^d_t(k^d)$.  Notice that due to equation \eqref{eq:bridge_rate} the particular structure of  the reference will lead in turn to transition rates for the conditional bridge processes with the following form:
\begin{eqnarray}
W_t(\*k,\*j|\*k_0,\*k_1) = \sum^D_{d=1} W_{t}^{d} (k^d , j^d | k^d_1)\prod_{l\neq d} \delta_{j^l,k^l}
\label{eq:trans_rate_indep}
\end{eqnarray}
Then again, by the Markovian structure in \eqref{eq:bridge_rate} the expression is independent of  the initial state $\*k_0$.  If we apply Eq.~\eqref{eq:marginalization_trick_rate} to Eq.~\eqref{eq:trans_rate_indep} we see that for  rates of the target process  only a single component $j^d$ of a vector $\*j$ will change. This  allows for  efficient simulations. That is, one has the form for the target:
\begin{equation}
W_t(\*k , \*j) = \sum^D_{d=1} W_{t}^d(k^d , \*j)\prod_{l \neq d}\delta_{k^l,j^l}
\end{equation}
where we have 
\begin{equation}
W_{t}^d (k^d , \*j) = \sum_{m =1}^S \pi_t(k_1^d =m  | \*j)\, W_{t}^d (k^d , j^d|k_1^d =m) 
\end{equation}
The required posterior probabilities $\pi_t(k_1^d = m | \*j)$ for the \textit{individual} coordinates of the end states $\*k_1$ 
given that the state $\*k_t = \*j$ are much more efficient
to be learned and approximated by neural networks compared to the full joint probability $\pi_t(\*k_1 | \*j)$ required in \eqref{eq:marginalization_trick_rate}. We now proceed to introduce a reference process with a  close form solution to the master equation.

\paragraph{Multivariate Random Telegraph process} We now introduce a simple reference process that leads to close form 
analytical solutions for $\tilde{q}$. This process is an $S+1$-state generalization of the Telegraph process for binary systems, typically used to model burst noise in semi-conductors or bit-flips in communication channels. We assume that the transition probability from all other states to a state $k$ is uniform and described by a rate function $\beta_t := R^d_{t}(k^d, j^d)$ for all $k$ and $j$, leading to the following  Master equation for $t > s$ :
\begin{equation}\label{eq:multivariate_telegraph}
    \partial_t \tilde{q}_{t | s} (n | m) = \beta(t)(1 - S \tilde{q}_{t | s} (n | m)),
\end{equation}
where for clarity we have omitted $d$. Here we focus on the the conditional distribution, as this equation corresponds to a master equation where we enforce  impose  condition $\tilde{q}_{s | s} (n | m) = \delta_{m,n}$.  In the expression \eqref{eq:multivariate_telegraph} we have use the fact that  $\sum_{l=1}^S q_{t | s} (l | m) = 1$. This linear, first-order differential equation is solved by
\begin{equation}\label{eq:telegram_propagator}
\tilde{q}_{t | s} (n | m) = 1/S + w_{t,s} \left(-1/S + \delta_{m,n}\right),
\end{equation}
where 
\begin{equation}
    w_{t,s} := \exp{\left(-S \int_{s}^{t} \beta(r) \, dr\right)}\,.
\end{equation}

In this work we will assume a constant rate function $\beta(t)=\beta$ with hyperparameter $\beta>0$. Using Eq.\eqref{eq:bridge_rate}, we obtain for dimension $d$ and $k \neq j$:
\begin{equation}
W^d_t(k^d=k , j^d=j | k_1) = 1 + \frac{w_{1,t} S}{1 - w_{1,t}} \delta_{k_1,k} + w_{1,t} \delta_{k_1,j},
\label{eq:conditional_telegram_rate}
\end{equation}
the equality follows from the binary nature of the Kronecker delta variables, i.e., $\delta_{k_1,k} \in \{0,1\}$. With the known expressions for the conditional rate in Eq.~\eqref{eq:conditional_telegram_rate}, we can now compute the averages over the posterior in Eq.~\eqref{eq:marginalization_trick_rate} to obtain $W_t^d$. This is derived from the posterior expectation as
\begin{align}
W_t^d(k^d , \mathbf{j}) &= \sum_{k_1=1}^S \pi_t(k_1 | \mathbf{j}) W^d_t(k^d , j^d| k_1) \notag \\
&= 1 + \frac{w_{1,t} S}{1 - w_{1,t}} \pi_t(k^d | \mathbf{j}) + w_{1,t} \pi_t(j^d | \mathbf{j}),
\label{eq:target_rate_from_posterior}
\end{align}
where the last equality is obtained by substituting Eq. \ref{eq:conditional_telegram_rate}.
To obtain the target rate $W_t$, one must learn to approximate the posterior in Eq. \ref{eq:poster_disc_discrete}, and from there, apply Eq. \ref{eq:target_rate_from_posterior}. Learning the posterior probabilities $q_t(\*k_0, \*k_1 | \*k)$ will involve drawing a large number of random samples $\*k_0, \*k_1, \*k_t$ from their joint probability according to the Markov bridge.

\subsection{Proofs}

\paragraph{Continuity Equation}
Here we show how one can construct the marginal vector field from an average of the conditional vector field:
\begin{align}\label{eq:marginalization_trick_continuity}
  \partial_t p_t(\mathbf{x}) 
  &\overset{(i)}{=} \int \partial_t p_t(\mathbf{x} | \mathbf{x}_0, \mathbf{x}_1) \mu(\mathbf{x}_0) \nu(\mathbf{x}_1) \, d\mathbf{x}_0 d\mathbf{x}_1 \\
  &\overset{(ii)}{=} - \int \nabla \cdot \big[ u_t(\mathbf{x} | \mathbf{x}_0, \mathbf{x}_1) p_t(\mathbf{x} | \mathbf{x}_0, \mathbf{x}_1) \big] \mu(\mathbf{x}_0) \nu(\mathbf{x}_1) \, d\mathbf{x}_0 d\mathbf{x}_1 \\
  &\overset{(iii)}{=} - \int \nabla \cdot \big[ u_t(\mathbf{x} | \mathbf{x}_0, \mathbf{x}_1) \rho(\mathbf{x}_0, \mathbf{x}_1 | \mathbf{x}) p_t(\mathbf{x}) \big] \, d\mathbf{x}_0 d\mathbf{x}_1 \\
  &\overset{(iv)}{=} - \nabla \cdot \Big[ p_t(\mathbf{x}) \int u_t(\mathbf{x} | \mathbf{x}_0, \mathbf{x}_1) \rho(\mathbf{x}_0, \mathbf{x}_1 | \mathbf{x}) \, d\mathbf{x}_0 d\mathbf{x}_1 \Big] \\
  &\overset{(v)}{=} - \nabla \cdot \big[ u_t(\mathbf{x}) p_t(\mathbf{x}) \big],
\end{align}
where we have use in (iii) bayes rule for the posterior:
\begin{equation}
    \rho_t(\mathbf{x}_0, \mathbf{x}_1 | \mathbf{x}) = \frac{p_t(\mathbf{x} | \mathbf{x}_0, \mathbf{x}_1)\mu(\mathbf{x}_0) \nu(\mathbf{x}_1)}{p_t(\mathbf{x})}.
\end{equation}
\paragraph{Master Equation} Here we show how one can construct the target rate from an average of the conditional rate:
\label{app:marginalization_trick_master}
\begin{align}
    \frac{\partial }{\partial t} q_t(\mathbf{k}) &= \frac{\partial }{\partial t} \sum_{\mathbf{k}_0, \mathbf{k}_1} q_t(\mathbf{k} | \mathbf{k}_0, \mathbf{k}_1) \mu(\mathbf{k}_0) \nu(\mathbf{k}_1)  \\
    &= \sum_{\mathbf{k}_0, \mathbf{k}_1} \frac{\partial }{\partial t} q_t(\mathbf{k} | \mathbf{k}_0, \mathbf{k}_1) \mu(\mathbf{k}_0) \nu(\mathbf{k}_1) \\
    &= \sum_{\mathbf{k}_0, \mathbf{k}_1} \sum_{\mathbf{j}} W_t(\mathbf{k} , \mathbf{j}| \mathbf{k}_1) q_t(\mathbf{j} | \mathbf{k}_0, \mathbf{k}_1) \mu(\mathbf{k}_0) \nu(\mathbf{k}_1) \\
    &\overset{(i)}{=} \sum_{\mathbf{j}} \left\{ \sum_{\mathbf{k}_1,\mathbf{k}_0} W_t(\mathbf{k} , \mathbf{j}| \mathbf{k}_1) \pi_t(\mathbf{k}_1, \mathbf{k}_0| \mathbf{j}) \right\} q_t(\mathbf{j})\\
    & = \sum_{\*j}{W}_t(\*k,\*j) \, {q}_t(\*j)
\end{align}
where we have use the posterior
\begin{equation}
    \pi_t(\mathbf{k}_0, \mathbf{k}_1 | \mathbf{j}) = \frac{q_t(\mathbf{j} | \mathbf{k}_0, \mathbf{k}_1) \mu(\mathbf{k}_0) \nu(\mathbf{k}_1)}{q_t(\mathbf{j})}.
    \label{eq:posterior_master}
\end{equation}

\paragraph{Conditional Rate}
We now obtain the rate of the reference process conditioned on the end points, that is a jump process bridge $P(\cdot|\mathbf{k}_1, \mathbf{k}_0)$. We omit the dimension index $d$ for clarity.
\begin{align}
    W_t(\mathbf{k},\mathbf{j}| \mathbf{k}_1) &= \lim_{\Delta t \to 0} \left[ \frac{\tilde{q}_{t+\Delta t| t}(\mathbf{k}|\mathbf{j}, \mathbf{k}_1) - \delta_{\mathbf{k},\mathbf{j}}}{\Delta t} \right] \\
    &= \lim_{\Delta t \to 0} \left[ \frac{\tilde{q}_{1,t+\Delta t,t}(\mathbf{k}_1, \mathbf{k}, \mathbf{j})}{\Delta t \tilde{q}_{1,t}(\mathbf{k}_1, \mathbf{j})} - \frac{\delta_{\mathbf{k},\mathbf{j}}}{\Delta t} \right] \\
    &= \lim_{\Delta t \to 0} \left[ \frac{\tilde{q}_{1|t+\Delta t}(\mathbf{k}_1|\mathbf{k}) \tilde{q}_{t+\Delta t|t}(\mathbf{k}|\mathbf{j}) \tilde{q}_t(\mathbf{j})}{\Delta t \tilde{q}_{1|t}(\mathbf{k}_1|\mathbf{j}) \tilde{q}_t(\mathbf{j})} - \frac{\tilde{q}_{1|t+\Delta t}(\mathbf{k}_1|\mathbf{k}) \delta_{\mathbf{k},\mathbf{j}}}{\tilde{q}_{1|t}(\mathbf{k}_1|\mathbf{j}) \Delta t} \right] \\
    &= \lim_{\Delta t \to 0} \left[ \frac{\tilde{q}_{1|t+\Delta t}(\mathbf{k}_1|\mathbf{k})}{\tilde{q}_{1|t}(\mathbf{k}_1|\mathbf{j})} \left( \frac{\tilde{q}_{t+\Delta t}(\mathbf{k}|\mathbf{j}) - \delta_{\mathbf{k},\mathbf{j}}}{\Delta t} \right) \right] \\
    &= \frac{\tilde{q}_{1|t}(\mathbf{k}_1|\mathbf{k})}{\tilde{q}_{1|t}(\mathbf{k}_1|\mathbf{j})} R_t(\mathbf{k};\mathbf{j})
\end{align}

\paragraph{Conditional Probability/ Markov Bridge}
\label{app:conditional_probability_bridge}
Here we obtain the expression for Eq.~\eqref{eq:discrete_bridge}, this equation holds for any Markov process
\begin{align*}
    p_t(\mathbf{x} | \mathbf{x}_0, \mathbf{x}_1) 
    &\overset{\text{(i)}}{=} \frac{p(\mathbf{x}_0, \*x_t = \mathbf{x},\*x_1)}{p(\mathbf{x}_0, \mathbf{x}_1)} \\ 
    &\overset{\text{(ii)}}{=} \frac{p(\mathbf{x}_1 | \mathbf{x}_t = \mathbf{x}) p(\mathbf{x}_t = \mathbf{x} | \mathbf{x}_0) p(\mathbf{x}_0)}
    {p(\mathbf{x}_0, \mathbf{x}_1)} \\ 
    &\overset{\text{(iii)}}{=} \frac{p_{1|t}(\mathbf{x} | \mathbf{x}) p_{t|0}(\mathbf{x} | \mathbf{x}_0) p(\mathbf{x}_0)}
    {p_{1|0}(\mathbf{x}_1| \mathbf{x}_0)p(\*x_0)} \\ 
    &= \frac{p_{1|t}(\mathbf{x}_1 | \mathbf{x}) p_{t|0}(\mathbf{x} | \mathbf{x}_0)}
    {p_{1|0}(\mathbf{x}_1 | \mathbf{x}_0)}
\end{align*}
Here as before one only requires the Markovianity assumption of $p$, which means that for $t > s$ one can write $p(\*x_t,x_s) = p_{t|s}(\*x_t|\*x_s)p(\*x_s)$ as well as Bayes' rule.

\section{Architecture}
\label{app:architecture}

\paragraph{Mode embeddings}
To effectively model the multimodal nature of the data, we embed independently each input mode, $x^d_t$ and $k^d_t$, into a high-dimensional vector space $\mathbb{R}^{h_{\tt emb}}$ using Multi-layer perceptrons ({\tt MLP}) consisting of two linear layers with a {\tt GELU} non-linear activation function in between. To process the flavor tokens $k^d_t$ we replace the first layer of the MLP with a learnable lookup table implemented in PyTorch via {\tt nn.Embedding}. The time variable $t\in [0,1]$ is encoded into $t_{\rm emb}\in \mathbb{R}^{h_{\tt emb}}$ using a Fourier feature ({\tt FF}) embedding of \cite{tancik2020fourier}. The resulting embeddings for each mode are then summed with the embedded time vector $t_{\rm emb}={\tt FF}(t)$ to form hidden kinematic and flavor representations: 
\begin{align}
\*x_t^\prime= {\tt MLP}(\*x_t) + t_{\rm emb}\,, \quad \*k_t^\prime= {\tt MLP}(\*k_t) + t_{\rm emb}\,.
\end{align}
These are subsequently fed into a time-dependent encoder. 

\begin{figure}[t]
    \centering
    \includegraphics[width=1.0\linewidth]{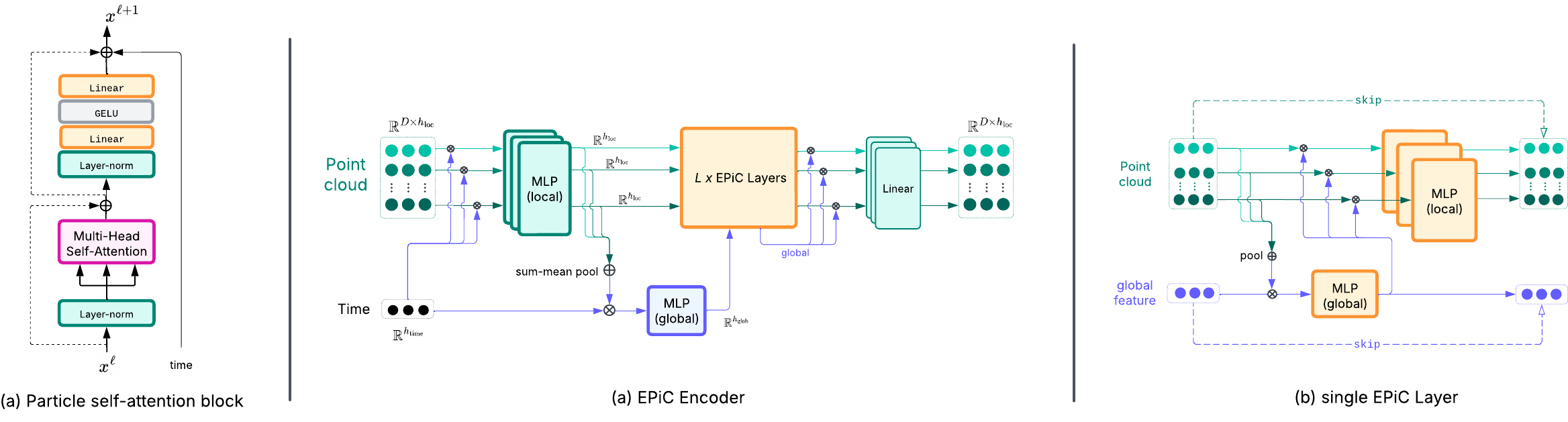}\\
    \caption{Detail of the particle self-attention block used for our MMF model and the EPiC encoder used for our baseline model.} \label{fig:particle_attn_epic}
\end{figure}

\paragraph{Multimodal particle transformer} 
Given that the dataset consists of particle clouds, where the ordering of constituents is unimportant, it is essential to approximate the generators of the dynamical process with a permutation-equivariant architecture. In this work we employ a particle transformer architecture, or {\it ParticleFormer} for short, as the core component of our multimodal encoder. The ParticleFormer processes the embedded time variable along with the embedded particle-level features with a stack of multi-head self-attention blocks. Details are shown in Fig. \ref{fig:particle_attn_epic} (a). We use a generic GPT-style self-attention block without causal masking and positional encoding to guarantee permutation equivariance. Particle transformers of this sort were first used for jet tagging on simulated data by \cite{Qu:2022mxj}, producing state-of-the-art results when compared to previous methods such as GANs and graph neural networks. 

Our multimodal encoder consists of two mode-specific encoders that feed into a {\it fused} encoder:
\begin{align}
F_{\rm kin} &=  {\tt ParticleFormer}_{\,L_1}\left(t_{\rm emb}, \, \*x_t^\prime\right)\,,\\
F_{\rm flav} &=  {\tt ParticleFormer}_{\,L_2}\left(t_{\rm emb}, \, \*k_t^\prime\right)\,,\\
F _{\rm fuse} &= {\tt ParticleFormer}_{\, L}\left(t_{\rm emb},  \,F_{\rm kin}\otimes F_{\rm flav}\right),
\end{align}
Here, $\otimes$ denotes feature concatenation, and the parameters $L$, $L_1$, and $L_2$ represent the number of self-attention layers in each encoder. This architecture provides flexibility, enabling the network to capture intra-modal correlations in the early mode-specific encoders, while cross-modal correlations between kinematics and flavor are learned in the subsequent fused encoder. Notice here that this particular choice of first encoding each modality and then fusing them is somewhat arbitrary. Other multimodal frameworks, like using cross-attention between modes could be implemented or added to our framework. Exploring other multimodal setups is left for future studies.  

The output state of the fused encoder is split into two equal-sized states $F_{\rm fuse} = H_{\rm kin}\otimes H_{\rm flav}$, which are then processed by the (continuous) regressor and the (discrete) classifier heads. Before passing these to each head, we add the mode-specific residual states and the time embedding:
\begin{align}
    H_{\rm kin}^\prime = H_{\rm kin} + F_{\rm kin} + t_{\rm emb}\,,\quad H_{\rm flav}^\prime = H_{\rm flav} + F_{\rm flav} + t_{\rm emb}\,.
\end{align}

The heads consist of two-layered MLPs with a ${\tt GELU}$ activation function in between that output the velocity field and the posterior classifier, respectively:

\begin{align}
    u_t^\theta \otimes h_t^\theta&= {\tt MLP}(H_{\rm kin}^\prime ) \otimes{\tt MLP}(H_{\rm flav}^\prime )\,.\label{eq:post_probs}
\end{align}

\paragraph{Uncertainty network} As discussed in the text, we promote the uncertainty weights of our multi-modal loss \eqref{eq:multimodal_loss} to time-dependent functions parametrized with an neural network. For this network we use a single Fourier feature layer with a 128-dimensional hidden state followed by a linear projection layer $(w^1_t,w^2_t)={\tt Linear}({\tt FF}(t))$. This module is only used during the training phase.

\section{Sampling algorithm details}
\label{app:sampling}
To generate the particle kinematics we directly solve the ODE \eqref{eq:ode} using any well-known integration method. For simplicity, we integrate using Euler's first-order method:
\begin{align}\label{eq:euler}
    x^d_{t+\Delta t} = x^d_{t} + u^{\theta,\,d}_t( x^d_{t}, k^d_{t})\, \Delta t 
\end{align}
where $\Delta t$ is a small time-step, and $u_t^{\theta,\,d}$ is the parametric velocity field for each particle. 

\paragraph{$\tau$-leaping}  To efficiently simulate the random telegraph process, we employ \textit{$\tau$-leaping}. Rather than resolving each individual transition sequentially, tau-leaping assumes the total number of per-particle transitions $\Delta n^d_m$ into the flavor token $m$, occurring within a small time window $\Delta t$, follows a {\it Poisson distribution}, 
\begin{equation}\label{eq:tauleap_step}
\Delta n^d_m \sim\, \text{Poisson}(\,{W}^{\theta,\,d}_t( k_{t+\Delta t}^d=m, k_t^d, x_t^d)\, \Delta t\,).
\end{equation}
This approximation holds under the assumption that individual jumps within $\Delta t$ occur independently and with probabilities proportional to $W^{\theta,\,d}_t \, \Delta t$. In this regime, the total number of jumps $\Delta n^d_j$  can be understood as arising from many independent Bernoulli trials, which, in the limit of small probabilities, naturally follows a Poisson distribution. By appropriately selecting $\Delta t$, tau-leaping provides a useful trade-off between accuracy and computational efficiency, enabling the effective simulation of discrete jumps without the need for explicitly resolving each transition at every infinitesimal time step. Explicitly, at each time step, the discrete flavor state for each particle is updated via 
\begin{equation}\label{eq:tauleap}
    k^d_{t+\Delta t} \, =\,  \left[\,k^d_t + \sum_{m=1}^S\,(m-k^d_t)\, \Delta n^d_j\,\right]\!\!\!\!\! \mod S \,.
\end{equation}
Here, we take the modulo of the vocabulary size $S$ to deal with cases where the updated state inside the bracket results in an integer outside of $\mathcal{F}=\{0,...,7\}$. An alternative is to clamp the output to the boundary values so they remain within $\mathcal{F}$, with the expense of biasing the generation towards these tokens (in our case these correspond to the photon and the anti-muon). 

The combined sampling procedure for the hybrid states consists of iteratively updating the continuous and discrete features using Euler steps \eqref{eq:euler} for the kinematics and tau-leaping steps \eqref{eq:tauleap} for the flavor tokens.

\paragraph{Temperature scaling}  
As discussed in Sec.~\ref{sec:aoj_dataset}, the training dataset exhibits a pronounced class imbalance, with jets containing much more photons and charged hadrons compared to neutral hadrons and leptons. Such imbalances are well known to hinder classification performance [\cite{johnson2019survey}]. A common way to alleviate their impact is to recalibrate the posterior probabilities \eqref{eq:post_probs} through \textit{temperature scaling} [\cite{guo2017calibration}]. Specifically, we introduce a temperature hyperparameter $T$ as
\begin{align}
\pi_t^\theta = {\tt softmax}\!\left(\frac{{\tt MLP}(H_{\rm flav}^\prime)}{T}\right)\,.
\end{align}
This rescaling of the logits is only applied during generation, leaving training unaffected. Larger values $T>1$ soften the logits, reducing the relative differences between classes and approaching a uniform distribution as $T \to \infty$. Conversely, smaller values $T<1$ sharpen the distribution, accentuating class differences and approaching a one-hot assignment in the limit $T \to 0$. Temperature scaling has also been widely used in natural language processing and classification tasks, where it is known to balance class probabilities and improve calibration. In our experiments, we investigate the impact of the temperature on generation quality. 

\paragraph{Particle multiplicities}  
A fundamental limitation of dynamics-based flow models comes from their inability to handle particle-clouds with varying number of particles. Despite masking zero-padded entries during training, the generation step requires explicit conditioning on the number of particles per jet. This constraint originates from the continuity equation~\eqref{eq:continuity_equation} governing the reference dynamics, which enforces particle-number conservation along each trajectory, thereby precluding the spontaneous creation or annihilation of particles during the evolution. To address this limitation and allow variable particle numbers within our generative framework we fit a $150$-dimensional categorical distribution to the empirical particle multiplicity distribution. During generation, for each jet we sample the particle multiplicity $N$ from this auxiliary model, and subsequently apply the corresponding mask ($N$ ones followed by $150-N$ zeros) to the initial source data.

\section{Experiment details}
\label{app:experiments}

\paragraph{EPiC-FM baseline}\label{app:flow_baseline}
We train the baseline on the same target AOJ datasets. However, since flow-matching can only handle continuous variables, the flavor token of each particle is one-hot encoded into unit vectors in $\mathbb{R}^{S}$ representing flavor assignment probabilities. For the target jets, these probabilities are concatenated with the particle kinematics, forming an augmented continuous feature vector $x_1^d\in\mathbb{R}^{3+S}$. We generate source point-clouds by drawing each point from a standard Gaussian over $\mathbb{R}^{3+S}$. After generation, to ensure each particle has a unique flavor assignment, we apply an $\tt argmax$ operation to the generated assignment probabilities and tokenize back to $\mathcal{F}$. This strategy has been successfully employed in previous works [\cite{Birk:2023efj, Araz:2024bom}] for jet and event datasets. For our experiments, we implement the EPiC-FM encoder described in \cite{Buhmann:2023zgc} depicted in Fig.~\ref{fig:particle_attn_epic} (b) with the following setup:  $n_{\rm layers}=16$ EPiC layers with $h_{\rm loc}=256$ and $h_{\rm glob}=16$ for the local and global hidden dimensions. The resulting model has around 5.9 million parameters. 
Optimization is perform with the {\tt Adam} algorithm [\cite{kingma2014adam}] with an effective batch size of 256 jets for a maximum of 1500 epochs. A cosine-annealing learning rate schedule is applied for the first 1000 epochs, decaying from $5\times 10^{-4}$ to $10^{-5}$, followed by 500 epochs at a fixed learning rate of $10^{-5}$. 

\paragraph{MMF training details} We set the gaussian smearing hyperparameter of the flow-matching component to $\sigma=10^{-5}$ and use a constant stochasticity parameter with $\beta=0.075$ for the multivariate telegraph process. During training, the time parameter $t$ is sampled uniformly from a slightly reduced unit interval $[\epsilon, 1-\epsilon]$ with $\epsilon=10^{-5}$, to prevent numerical instabilities caused at the time boundaries $t=0,1$ for the MJB model. We parameterize the combined generators $u_t^\theta \otimes \pi_t^\theta$ using the multimodal architecture introduced in Sec.~\ref{sec:architecture} with a \textit{mid-fusion} setup $(L_1,L_2,L)=(5,5,6)$. This configuration balances the intra-modal and cross-modal correlations between kinematics and flavor tokens in separate sub-modules with similar sizes. We fix the number of attention heads and hidden dimensions to $n_{\rm heads}=4$, $n_{\rm embd}=256$, and $n_{\rm inner}=512$, resulting in a model with approximately 5.6 million trainable parameters. Training is performed with the uncertainty weighted loss of Eq.~\eqref{eq:multimodal_loss}. We use an uncertainty network with a 128-dimensional hidden state. Optimization is perform with the {\tt Adam} algorithm [\cite{kingma2014adam}] with an effective batch size of 256 jets for a maximum of 1500 epochs. A cosine-annealing learning rate schedule is applied for the first 1000 epochs, decaying from $5\times 10^{-4}$ to $10^{-5}$, followed by 500 epochs at a fixed learning rate of $10^{-5}$. 

The best models are chosen according to the lowest validation loss. All experiments are run on 16 NVIDIA A100 GPUs.

\end{document}